\documentclass[prd,twocolumn,reprint,preprintnumbers,nofootinbib,superscriptaddress]{revtex4-2}

\input{header.tex}
\begin{document}
\raggedbottom

\preprint{CERN-TH-2026-226}

\title{SHiP as a (post-)discovery machine: identifying the diphoton signals' origin}

\author{Matei~Climescu}
\email{matclim@cern.ch}
\affiliation{Ghent University, Ghent, Belgium}
\author{Malte~Fogde~Mikkelsen}
\email{maltefogde@gmail.com}
\affiliation{Department of Physics and Astronomy, Aarhus University, DK-8000 Aarhus C, Denmark}
\author{Maksym~Ovchynnikov}
\email{maksym.ovchynnikov@cern.ch}
\affiliation{Theoretical Physics Department, CERN, 1211 Geneva 23, Switzerland}

\date{\today}

\begin{abstract}
The upcoming SHiP experiment may probe interaction strengths of decaying feebly coupled particles several orders of magnitude below existing limits. In the event of a discovery, it may also reveal the nature of the new particle. This requires identifying its exact production mechanism, which SHiP does not observe directly. We explore whether this mechanism can be inferred from the kinematics of its visible decays. As a benchmark, we consider an axion-like particle (ALP) with a dominant diphoton decay, which may occur for couplings to the $U(1)_Y$ and $SU(2)_L$ gauge fields. For a known ALP mass, the observed energy and angular distributions depend on the relative magnitude and sign of the couplings and on the unknown lifetime, which we profile independently under each hypothesis. We first determine how the ALP energy and its longitudinal and transverse decay positions distinguish the interactions, assuming perfect reconstruction and accounting for production, propagation, decay, and geometric acceptance. We then include the detector response and reconstruct the diphoton decays using a full simulation of the SHiP electromagnetic calorimeter. Only $2$--$4$ reconstructed diphoton events are needed to distinguish the two pure $U(1)_{Y}$ and $SU(2)_{L}$ interactions for ALP masses between $0.2$ and $1~\mathrm{GeV}$. For coupling admixtures, the study assuming perfect reconstruction indicates requirements of several tens to $\mathcal{O}(10^2)$ events in favorable regions.
\end{abstract}

\maketitle

\begin{figure*}[t!]
  \centering
  \includegraphics[width=0.49\textwidth]{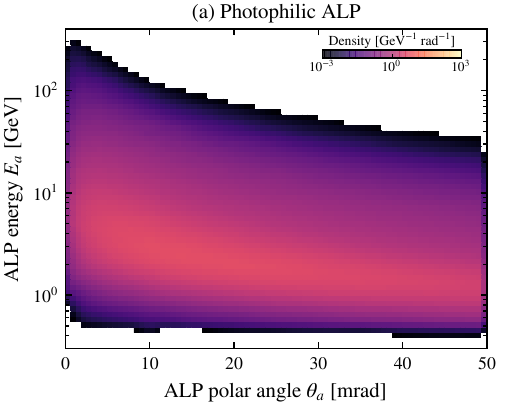}\hfill
  \includegraphics[width=0.49\textwidth]{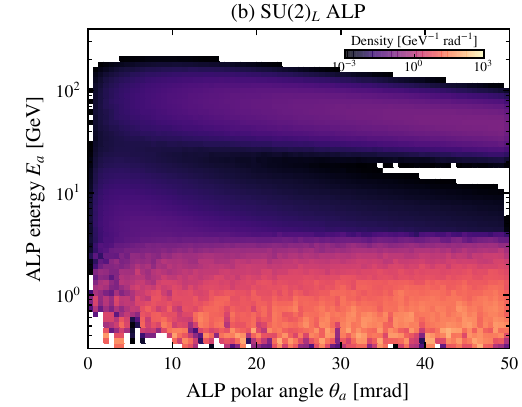}
  \caption{Joint density $\rho_H^{\rm prod}$ in the ALP energy and polar angle, defined in \cref{eq:production-density}, at $m_a=0.3~\gev$. Panel (a) combines Primakoff production from primary and cascade photons. Panel (b) combines decays of $B$ mesons and charged kaons with photon production induced by the $\SUtwo$ operator. Each panel is normalized over the displayed range, $\theta_a<0.05~\mathrm{rad}$.}
  \label{fig:production-distributions}
\end{figure*}

\section{Introduction}
\label{sec:introduction}

SHiP will search for decays of feebly interacting particles produced by $400~\gev$ protons striking a thick target~\cite{Aberle:2839677,SHiP:2025ows}. In favorable regions of parameter space, it can observe tens or hundreds of decays of a new particle. Such a discovery would make the identification of the particle and its interactions a central experimental task. The information available in the observed decays therefore determines how far SHiP can progress toward understanding the underlying physics.

The particle content and relative probabilities of different decay channels can help identify the new particle~\cite{Mikulenko:2023iqq}. For hadronic decays, these probabilities depend on the particle's interactions and quantum numbers, and final states containing several charged and neutral particles can dominate the observable signal~\cite{Kryshtal:2026bhq}. Angular correlations in the decay kinematics can also help determine the spin of hidden sector particles~\cite{AristizabalSierra:2026jsm} and probe their oscillations~\cite{Tastet:2019nqj}.

An important ambiguity remains when different interactions produce the same particle with the same visible decay. In the baseline SHiP setup, production occurs inside the thick target and cannot be observed directly. Models in which several new particles are produced together may offer additional information through multiple displaced decays~\cite{DallaValleGarcia:2025aeq,Bernreuther:2025xqk}. Here we focus on production channels containing one long-lived particle per collision, for which the production interaction must be inferred from the kinematics of its decay.

Axion-like particles (ALPs) with a dominant diphoton decay provide a clean example. The reach for a photophilic ALP at proton beam dumps has been studied extensively. It can be produced from primary photons and from the electromagnetic cascade initiated in the target, giving SHiP strong discovery potential~\cite{Dobrich:2015jyk,Dobrich:2019dxc,Patrone:2025fwk,Climescu:2026esp,Ovchynnikov:2025gpx}. Other ALP interactions, including electroweak and gluonic operators, can lead to the same diphoton final state while changing the production mechanism~\cite{Brivio:2017ije,Bauer:2021mvw,Ovchynnikov:2025gpx}. Simulation-based inference has previously been used to reconstruct the ALP mass and lifetime under an assumed $B\to Ka$ production mechanism~\cite{Morandini:2023pwj} and to search for excesses in model parameter space~\cite{Chathirathas:2024mep}. We take the inference one step further: the production interaction itself is unknown, and we determine it while profiling the lifetime independently under each hypothesis.

We investigate this question by comparing a photophilic ALP, produced by primary and cascade photons, with an ALP coupled to $\mathrm{SU}(2)_L$. The latter is produced through decays of $B$ mesons and charged kaons, together with photon sources induced by the same operator. Previous studies establish the SHiP reach for these interactions~\cite{Ovchynnikov:2023cry,Blinov:2024pza,Patrone:2025fwk,Ferber:2022rsf}. The ALP mass can be reconstructed from the two photons, while its unknown lifetime changes which ALPs reach and decay inside the detector. Consequently, a change in lifetime can imitate a change in the production interaction. We ask whether the ALP energy $E_a$ and the longitudinal and transverse decay positions $z$ and $\rperp$ resolve this ambiguity and can establish whether both interactions contribute.

The analysis proceeds in two complementary parts. We first determine the number of events needed to distinguish the interactions assuming perfect reconstruction, while accounting for production, propagation, decay, and geometric acceptance. We then test this capability with a full simulation of the SHiP electromagnetic calorimeter (ECAL), including photon showers, digitization, clustering, and reconstruction of the displaced decays~\cite{Bonivento:2018eqn,Albanese:2948477,Climescu:2025kdj}.

\section{Phenomenology of the two ALP models at SHiP}
\label{sec:phenomenology}

\begin{figure*}[t!]
  \centering
  \includegraphics[width=\textwidth]{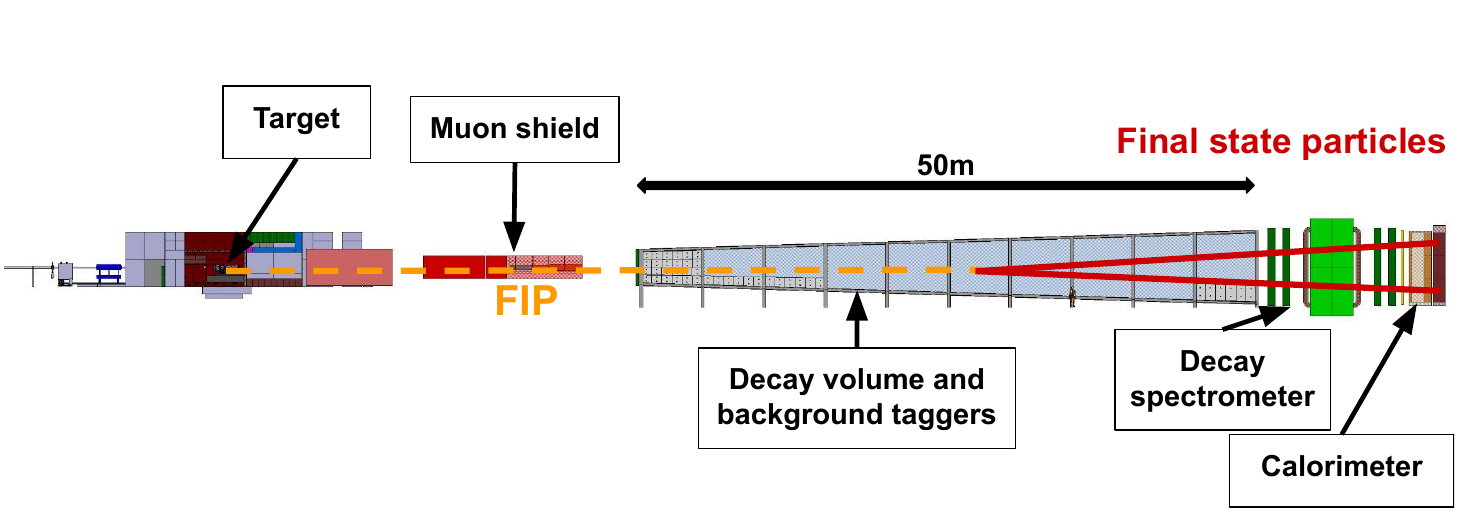}
  \caption{SHiP layout, with the photons from an ALP decay shown in red. The ECAL position and photon intersections used in the calculation are defined in \cref{eq:ecal-plane,eq:photon-intersection}. The event selections are defined in \cref{sec:geometry,sec:reco-samples}.}
  \label{fig:ship-schematic}
\end{figure*}

We define the two ALP models and describe their diphoton decays and production at SHiP. We then compare the resulting energy and angular distributions to identify the kinematic differences between the models.

\subsection{Model definition}
\label{sec:interactions}

The photophilic model is defined at low energy by
\begin{equation}
  \mathcal{L}_{\gamma}
  = \frac{\gagg}{4}aF_{\mu\nu}\widetilde F^{\mu\nu},
  \label{eq:Lgamma}
\end{equation}
where $F_{\mu\nu}$ is the electromagnetic field strength tensor and $\widetilde F^{\mu\nu}=\epsilon^{\mu\nu\alpha\beta}F_{\alpha\beta}/2$. Its diphoton width is
\begin{equation}
  \Gamma_{\gamma\gamma}^{(\gamma)}
  = \frac{\gagg^2 m_a^3}{64\pi}.
  \label{eq:width-photon}
\end{equation}

The electroweak model contains a single $\SUtwo$ operator at the matching scale,
\begin{equation}
 \begin{aligned}
  \mathcal{L}_{W}
  &= C_WaW^I_{\mu\nu}\widetilde W^{I\mu\nu},\\
  C_W&=\frac{\alpha_2}{4\pi}g_W,
  \qquad \frac{c_W}{f_a}=-C_W,
 \end{aligned}
  \label{eq:LW}
\end{equation}
with $\alpha_2=g_2^2/(4\pi)=\alpha_{\rm em}/\sin^2\theta_W$. Here $I=1,2,3$ labels the three $\mathrm{SU}(2)_L$ isospin generators. The final relation converts to the convention of Ref.~\cite{Gavela:2019wzg}, which writes the operator with coefficient $-c_W/f_a$. Electroweak symmetry breaking induces a photon coupling in the conventions of \cref{eq:Lgamma}~\cite{Gavela:2019wzg,Alonso-Alvarez:2018irt},
\begin{equation}
  \gagg^{(W)}=\frac{\alpha_{\rm em}}{\pi}g_W
  \simeq 2.32\times10^{-3}g_W,
  \label{eq:induced-photon}
\end{equation}
and hence
\begin{equation}
  \Gamma_{\gamma\gamma}^{(W)}
  =\frac{\alpha_{\rm em}^2m_a^3g_W^2}{64\pi^3}.
  \label{eq:width-W}
\end{equation}

Both models decay predominantly to two photons over the mass range considered here. Dalitz decays through an off-shell photon and other subleading final states contribute at the percent level in the photophilic case~\cite{EscuderoAbenza:2025tsi}. Radiatively induced leptonic modes remain subleading for the pure $c_W$ model~\cite{Gavela:2019wzg}. At $m_a=0.3~\gev$, the calculated widths give $\br(a\to\gamma\gamma)\simeq98$--$99\%$ in both models, with a difference below 0.5\%. The curves of the accepted yield below thus use the semianalytic convention $\br(a\to\gamma\gamma)=1$.

A general ALP can contain both interactions. Their coefficients in the Lagrangian are
\begin{equation}
 C_\gamma^{\rm dir}=\frac{g_{a\gamma\gamma}^{\rm dir}}{4},
 \qquad
 C_W=\frac{\alpha_2}{4\pi}g_W.
 \label{eq:mixed-operator-coefficients}
\end{equation}
After electroweak symmetry breaking, the coefficient of $aF_{\mu\nu}\widetilde F^{\mu\nu}$ is
\begin{equation}
 C_\gamma^{\rm tot}
 =C_\gamma^{\rm dir}+\sin^2\theta_W C_W.
 \label{eq:mixed-photon-coupling}
\end{equation}
The two terms add coherently, so their relative sign affects photon production and the diphoton width. We parameterize the size of the electroweak operator by
\begin{equation}
 \xi=\frac{|C_W|}{|C_\gamma^{\rm dir}|+|C_W|}.
 \label{eq:mixed-xi}
\end{equation}
The limits $\xi=0$ and $1$ give the pure photophilic and pure electroweak operators. Defining $\Lambda=|C_\gamma^{\rm dir}|+|C_W|$, we write
\begin{equation}
 C_W=\Lambda\xi,
 \qquad
 C_\gamma^{\rm dir}=s\Lambda(1-\xi),
 \qquad
 s=\pm1,
 \label{eq:mixed-sign}
\end{equation}
so that $C_\gamma^{\rm tot}/\Lambda=s(1-\xi)+\sin^2\theta_W\xi$. The choices $s=+1$ and $-1$ define constructive and destructive interference. The production fractions are nonlinear functions of $\xi$ because the electroweak interaction contributes both flavor-changing and photon production.

\subsection{Production mechanisms and their hierarchy}
\label{sec:production}
\begin{figure*}[t!]
  \centering
  \includegraphics[width=0.49\textwidth]{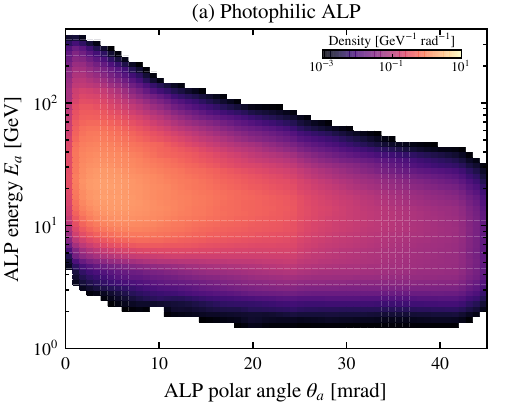}\hfill
  \includegraphics[width=0.49\textwidth]{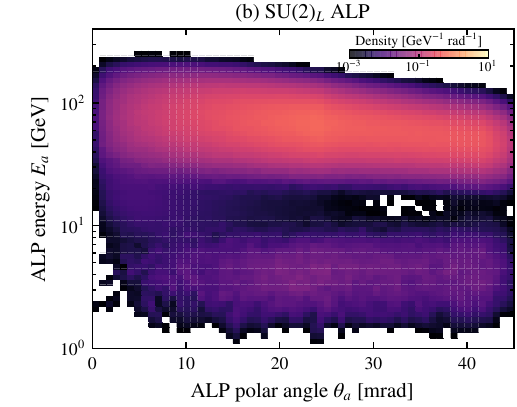}
  \caption{Joint density $\rho_H^{\rm traj}$ in the ALP energy and polar angle, defined in \cref{eq:trajectory-density}, at $m_a=1~\gev$, for trajectories that intersect the SHiP fiducial volume. Panel (a) combines primary and cascade Primakoff production; panel (b) combines $B$ decays and photon production induced by the $\SUtwo$ operator. Charged-kaon decays are forbidden at this mass. The distributions are normalized after the trajectory requirement, before applying the decay probability.}
  \label{fig:source-distributions}
\end{figure*}

\begin{figure*}[t!]
  \centering
  \includegraphics[width=0.49\textwidth]{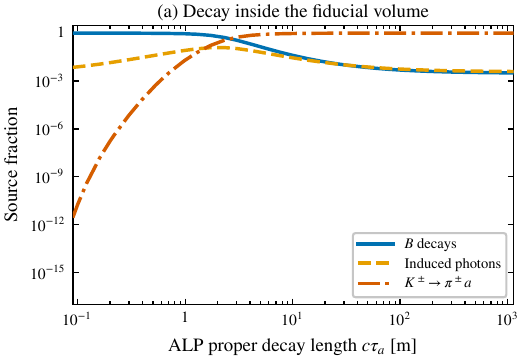}\hfill
  \includegraphics[width=0.49\textwidth]{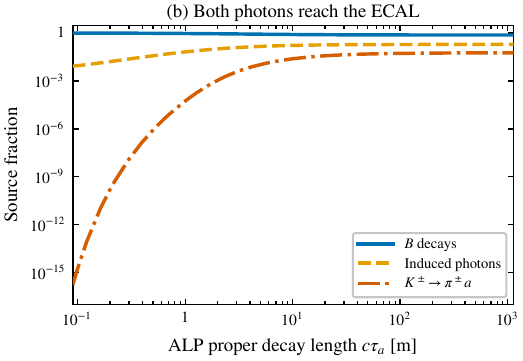}
  \caption{Contributions of $B$ decays, Primakoff production, and charged-kaon decays to the $\SUtwo$ signal at $m_a=0.3~\gev$. Each curve shows $F_s=\mu_s/\sum_{s'}\mu_{s'}$, with $\mu_s$ the yield from source $s$ in \cref{eq:event-yield}. Panel (a) counts decays inside the vessel, setting $\epsilon_{H,s}=1$; panel (b) includes the requirement that both photons reach the ECAL. The Primakoff contribution combines primary and cascade photons.}
  \label{fig:kaon-source}
\end{figure*}

Photophilic ALPs can be produced through decays of neutral mesons, photon fusion, and Primakoff scattering initiated by photons from the proton interaction in the target~\cite{Dobrich:2015jyk,Dobrich:2019dxc,Jerhot:2022chi}. The benchmark studied here contains the primary Primakoff source and the additional Primakoff source from secondary photons in the target shower~\cite{Blinov:2024pza,Patrone:2025fwk}. Decays of neutral mesons and photon fusion lie outside this production benchmark. The two included photon components differ in energy and angle and are propagated separately through the decay and geometric acceptance.

For the pure $\SUtwo$ model, flavor-changing transitions induced by loops, $b\to sa$ and $b\to da$, produce ALPs through
\begin{equation}
  B\to X_s a,
  \qquad
  B\to \pi a,
  \label{eq:B-modes}
\end{equation}
Here, $X_s$ denotes the inclusive sum over strange hadronic final states; we include $K$, $K^*$, $K_0^*$, $K_1$, and $K_2^*$. These channels dominate, while the pion channel, suppressed by the CKM matrix, extends the upper mass range~\cite{Gavela:2019wzg,Bauer:2021mvw}. Their relative rates follow the exclusive ratios calculated for the scalar portal: after neglecting light quark masses, the derivative flavor-changing neutral-current interaction reduces through the quark equations of motion to the same leading chiral hadronic current. The derivation and production normalizations are given in Appendix~\ref{sup:production}.

The same flavor-changing interaction allows $K^\pm\to\pi^\pm a$ below $m_K-m_\pi$. We take the energy and angular spectrum of charged kaons from Ref.~\cite{Gorbunov:2020rjx} and normalize it to $0.36$ decays in flight per proton on target, summed over both charges. The branching fraction uses the complete quark loop amplitude~\cite{Gavela:2019wzg}. In the limit of a light ALP, $m_a\ll m_K-m_\pi$,
\begin{equation}
 \frac{\br(K^\pm\to\pi^\pm a)}{g_W^2}
 =9.00\times10^{-4}~\gev^2.
 \label{eq:kaon-branching-coefficient}
\end{equation}
For $m_a=0.3~\gev$ and $\theta_a<0.05~\mathrm{rad}$, the yields from charged kaons, $B$ mesons, and Primakoff production are approximately in the ratio $66:19:1$. The Primakoff contribution includes primary and cascade photons from the induced coupling in \cref{eq:induced-photon}. The total yield is approximately $10^{-14}$ ALPs per proton on target at $g_W=10^{-4}~\gev^{-1}$ and scales as $g_W^2$.

\subsection{Energy and angle distributions at production}
\label{sec:kinematic-imprint}

The parent particles determine the ALP energy and direction at the target. We calculate the production distributions with \textsc{SensCalc}~\cite{Ovchynnikov:2023cry}, a semianalytic framework for predicting signal yields at different experiments. \textsc{EventCalc-SHiP}~\cite{EventCalc} is an event generator that uses the production and decay inputs from \textsc{SensCalc} to generate individual events specifically for SHiP. The primary and cascade photon spectra follow the treatment of Ref.~\cite{Climescu:2026esp}; the spectra of $B$ mesons and charged kaons simulated for the thick SHiP target follow Refs.~\cite{CERN-SHiP-NOTE-2015-009,Gorbunov:2020rjx}, respectively.

For hypothesis $H\in\{\gamma,W\}$, let $f_{H,s}(\theta_a,E_a)$ be the normalized production distribution from source $s$, and $P_{H,s,\rm prod}$ its production probability per proton on target. We combine the sources with these probabilities and normalize the plotted density over the displayed region $\Omega$ in energy and angle:
\begin{align}
 \Phi_H(\theta_a,E_a)
 &=\sum_s P_{H,s,\rm prod}f_{H,s}(\theta_a,E_a),\nonumber\\
 \rho_H^{\rm prod}(\theta_a,E_a)
 &=\frac{\Phi_H(\theta_a,E_a)}
 {\int_\Omega\diff\theta_a\,\diff E_a\,\Phi_H(\theta_a,E_a)}.
 \label{eq:production-density}
\end{align}

\Cref{fig:production-distributions} shows the ALP distributions before propagation through SHiP. Photophilic production is concentrated at small polar angles and spans a broad energy range. In the electroweak model, decays of charged kaons produce the soft component at broad angles, while $B$ decays populate the harder component. The photon sources induced by the electroweak operator add the forward pattern also present in the photophilic model.

The ALP boost influences its likelihood to reach and decay inside the vessel. Its polar angle determines the transverse displacement of the parent trajectory and the probability that both photons remain inside the aperture of the ECAL, the detector subsystem used here to reconstruct the diphoton decay.

\section{Semianalytic calculation of the SHiP diphoton signal}
\label{sec:event-samples}

In this section, we calculate the diphoton yield semianalytically using generated decay kinematics. The analysis including detector response and event reconstruction is presented in \cref{sec:ecal-reco}.

\subsection{SHiP geometry and geometric acceptance}
\label{sec:geometry}

The SHiP detector for hidden sector decays begins roughly $33~\m$ downstream of the target and contains a decay volume about $50~\m$ long~\cite{Aberle:2839677,SHiP:2025ows}. The longitudinal coordinate $z$ is measured from the target, and $z_{\min}$ denotes the entrance to the decay volume. The analysis uses the corresponding \textsc{EventCalc-SHiP} fiducial geometry and approximates the ECAL as a $4~\m\times6~\m$ plane centered on the beam axis at
\begin{equation}
  z_{\rm ECAL}=95~\m.
  \label{eq:ecal-plane}
\end{equation}
For a decay at $(x_d,y_d,z_d)$, photon $i$ intersects this plane at
\begin{align}
 x_i^{\rm ECAL}&=x_d+(z_{\rm ECAL}-z_d)\frac{p_{x,i}}{p_{z,i}},\\
 y_i^{\rm ECAL}&=y_d+(z_{\rm ECAL}-z_d)\frac{p_{y,i}}{p_{z,i}}.
 \label{eq:photon-intersection}
\end{align}
For a decay inside the fiducial volume, we call the event accepted if both photons move downstream and satisfy $|x_i^{\rm ECAL}|\leq2~\m$ and $|y_i^{\rm ECAL}|\leq3~\m$.

For an ALP produced at the target, its transverse decay coordinate and polar angle satisfy
\begin{equation}
  \rperp=z\tan\theta_a,
  \qquad
  \theta_a\simeq\arctan\!\left(\frac{\rperp}{z}\right).
  \label{eq:r-angle}
\end{equation}
The generated decay coordinates therefore retain information about the production angle.

We use the four event definitions in \cref{tab:event-levels} throughout.

\begin{table}[t]
\centering
\caption{Definitions of the four event samples used throughout the analysis. The geometry in \cref{sec:geometry} defines the fiducial and accepted samples; \cref{sec:reco-samples} defines the reconstructed sample.}
\label{tab:event-levels}
\begin{ruledtabular}
\begin{tabular}{ll}
Term & \parbox[t]{0.62\columnwidth}{Definition} \\
\hline
Produced & \parbox[t]{0.62\columnwidth}{ALP created in the target.} \\
Fiducial & \parbox[t]{0.62\columnwidth}{Produced ALP decays inside the hidden sector volume.} \\
Accepted & \parbox[t]{0.62\columnwidth}{Fiducial decay with both generated photons intersecting the simplified ECAL aperture.} \\
Reconstructed & \parbox[t]{0.62\columnwidth}{Two reconstructed photons pass chosen analysis selections.} \\
\end{tabular}
\end{ruledtabular}
\end{table}

\subsection{Determination of the accepted event yield}
\label{sec:eventcalc}

We denote the coupling by $g_H$, with $g_\gamma=\gagg$ for the photophilic model and $g_W$ defined in \cref{eq:LW}. The tabulated distributions in angle and energy, together with the production probabilities, are incorporated in \textsc{EventCalc-SHiP}. Each distribution keeps its normalization over the full physical kinematic domain. An input truncated in polar angle therefore retains the fraction inside that angular range. The accepted event yield follows the semianalytic prescription used for the cascade ALP study,
\begin{equation}
\begin{aligned}
  \mu_H^{\rm acc}(m_a,g_H)
  &=N_{\pot}\br_H(a\to\gamma\gamma)
    \sum_s P_{H,s,\rm prod}(m_a,g_H)\\
  &\quad\times\int\diff\theta_a\,\diff E_a\,\diff z\,
    f_{H,s}(\theta_a,E_a)\\
  &\quad\times\frac{\Delta\phi(\theta_a,z)}{2\pi}
    \frac{\diff P_{\rm dec}}{\diff z}
    \epsilon_{H,s}(\theta_a,E_a,z).
\end{aligned}
\label{eq:event-yield}
\end{equation}
where $\Delta\phi(\theta_a,z)$ is the azimuthal range for which the parent trajectory lies inside the transverse boundaries of the decay vessel. The factor $\epsilon_{H,s}$ averages the probability that both photons hit the ECAL over the allowed azimuthal interval of the parent and the isotropic diphoton decay angles. The longitudinal decay law is
\begin{equation}
 \frac{\diff P_{\rm dec}}{\diff z}
 =\frac{1}{\ell_z}\exp\!\left(-\frac{z}{\ell_z}\right),
 \qquad
 \ell_z=c\tau_a\frac{p_a}{m_a}\cos\theta_a.
 \label{eq:decay-law}
\end{equation}

The primary photon, cascade photon, $B$ decay, and charged kaon sources are evaluated separately because their spectra and ECAL acceptances differ. For electroweak ALPs, the probabilities for photon production follow the induced coupling in \cref{eq:induced-photon}; charged kaons contribute below $m_K-m_\pi$. Each source enters with its absolute \textsc{SensCalc} production yield. We add the propagated contributions before normalizing the combined distribution. Production probabilities scale as $g_H^2$ for either hypothesis containing one interaction. The planned exposure over 15 years of operation is
\begin{equation}
  N_{\pot}=6\times10^{20}.
  \label{eq:exposure}
\end{equation}

\textsc{EventCalc-SHiP} places the ALP production point at the target and uses the geometry in \cref{sec:geometry}.

Before generating decays, we select ALP trajectories that intersect the fiducial volume. Let $a_{\rm geom}(\theta_a)$ be the fraction of azimuthal directions for which this occurs. The resulting density is
\begin{equation}
 \rho_H^{\rm traj}(\theta_a,E_a)
 =\frac{a_{\rm geom}(\theta_a)\Phi_H(\theta_a,E_a)}
 {\int_\Omega\diff\theta_a\,\diff E_a\,
 a_{\rm geom}(\theta_a)\Phi_H(\theta_a,E_a)}.
 \label{eq:trajectory-density}
\end{equation}
\Cref{fig:source-distributions} shows this density at $m_a=1~\gev$. Photophilic production remains softer than the electroweak distribution, which is dominated by $B$ decays.

The requirement that both photons reach the ECAL acts most strongly on soft ALPs, whose photons have larger opening angles. At low mass, decays of charged kaons dominate the electroweak ALPs that decay inside the vessel over much of the lifetime range. Their contribution becomes small once both photons must reach the ECAL, leaving $B$ decays and induced photon production as the main sources, as shown in \cref{fig:kaon-source}.

\section{Statistical method for comparing production interactions}
\label{sec:statistics}

The statistical question is whether $N$ accepted diphoton events are more compatible with photophilic production or with $\SUtwo$ production. In this section, we assume that the energies and trajectories of the two photons are measured with perfect accuracy, allowing exact reconstruction of the ALP energy, invariant mass, and decay position. Finite measurement accuracy and its effects on diphoton reconstruction are treated in \cref{sec:ecal-reco}. We fit the unknown lifetime independently under each hypothesis and compare the normalized kinematic distributions. We first define the comparison between pure interactions and then extend the test to mixtures.

\subsection{Choosing lifetimes for the model comparison}
\label{sec:lifetime-domains}

The ALP lifetime controls whether the particle reaches the decay volume and decays inside it. SHiP therefore probes a finite range of lifetimes, which depends strongly on the ALP mass and production interaction. At fixed mass, the proper decay length scales with the coupling as $c\tau_a(m_a,g_H)\propto g_H^{-2}$. For small couplings such that $c\tau_a(m_a,g_H)\langle\gamma_a\rangle\gg100~\m$, too few ALPs decay over the distances relevant to SHiP. Here, $\langle\gamma_a\rangle=\langle E_a\rangle/m_a$ is the mean Lorentz factor, so $c\tau_a\langle\gamma_a\rangle$ is the characteristic decay length for relativistic ALPs. For large couplings such that $c\tau_a(m_a,g_H)\langle\gamma_a\rangle\ll z_{\min}$, most ALPs decay before reaching the decay volume.

For the exposure in \cref{eq:exposure} and negligible background, we define the sensitivity boundary by a mean of 2.3 accepted decays,
\begin{equation}
  \mu_H^{\rm acc}(m_a,g_H)=2.3.
  \label{eq:sensitivity}
\end{equation}
This corresponds to a 90\% probability of observing at least one event. As an example of the lifetime coverage, at $m_a=0.3~\gev$ this criterion reaches proper decay lengths of approximately $575~\m$ for the photophilic model and $1310~\m$ for the $\SUtwo$ model, before imposing existing constraints. To illustrate the two production spectra in the long lifetime regime at this mass, we use a common benchmark $c\tau_a\simeq90~\m$ in \cref{fig:energy-lifetime-degeneracy}. This value lies within the sensitivity range of both models and remains allowed by existing constraints.

Distinguishing the production interactions can require more events than observing a signal. For a comparison using $N$ accepted decays, we therefore retain lifetimes for which the hypothesis predicts a mean accepted yield of at least $N$,
\begin{equation}
 \mathcal{T}_H(m_a,N)=
 \left\{c\tau>0\mid \mu_H^{\rm acc}(m_a,c\tau)\geq N\right\}
 \setminus\mathcal{E}_H(m_a),
 \label{eq:lifetime-domain}
\end{equation}
where $\mathcal{E}_H$ denotes the lifetime intervals excluded by existing constraints. For each hypothesis, this domain restricts both the lifetimes used to test the discrimination and those allowed in the fit. The restriction concerns the expected yield; Poisson fluctuations can still produce $N$ observed events when the mean is below $N$. Disconnected intervals are treated separately. The quoted numbers of events required apply to the adopted exposure and geometric selection.

\subsection{Distributions of energy and decay position}
\label{sec:probability-model}

For model $H$ and lifetime $c\tau_H$, the probability that an accepted event lies in energy bin $i$ is
\begin{equation}
 p_{H,i}(c\tau_H)=
 \frac{\sum_{k\in i}w_{H,k}(c\tau_H)}
      {\sum_kw_{H,k}(c\tau_H)}.
 \label{eq:energy-probability}
\end{equation}
The two models use the same energy bins at each mass. The bins are merged until every probability entering the lifetime fit is supported by an effective simulated sample of at least 100, where $N_{\rm eff}=(\sum_k w_k)^2/\sum_k w_k^2$ for the weighted events in that bin. Appendix~\ref{sup:pure-likelihood} gives the bin construction and its numerical dependence.

The spatial vector is
\begin{equation}
 \boldsymbol f=
 \begin{pmatrix}z\\ \rperp\end{pmatrix},
 \qquad
 \rperp=\sqrt{x_d^2+y_d^2}.
 \label{eq:spatial-vector}
\end{equation}
For accepted events in energy bin $i$, the simulation gives the conditional mean $\boldsymbol\mu_{H,i}$ and covariance $\boldsymbol\Sigma_{H,i}$. For $N$ simulated events with energy bins $\boldsymbol i=(i_1,\ldots,i_N)$, the analysis retains those bin labels and the mean $\overline{\boldsymbol f}$ of their spatial coordinates. The corresponding likelihood is
\begin{multline}
 \log\mathcal{L}_H(c\tau_H)=
 \sum_{k=1}^N\log p_{H,i_k}(c\tau_H)\\
 +\log\mathcal{N}_2\!\left(
 \overline{\boldsymbol f}\mid
 \overline{\boldsymbol\mu}_H(c\tau_H\mid\boldsymbol i),
 \boldsymbol V_H(c\tau_H\mid\boldsymbol i)
 \right).
 \label{eq:likelihood}
\end{multline}
Conditioning the Gaussian term on the observed energy bins retains correlations between energy and the mean spatial coordinates. Pseudoexperiments draw the spatial means from this same Gaussian distribution, so the quoted thresholds for few events depend on this approximation. Appendix~\ref{sup:pure-likelihood} gives the conditional moments and simulation sizes.

\subsection{Likelihood comparison and required number of events}
\label{sec:test-statistic}

For each hypothesis, we maximize the likelihood over the domain in \cref{eq:lifetime-domain},
\begin{equation}
 \widehat\ell_H(N)=\max_{c\tau_H\in\mathcal{T}_H(m_a,N)}\log\mathcal{L}_H(c\tau_H),
 \label{eq:profiled-likelihood}
\end{equation}
and the test statistic is
\begin{equation}
 T=2\left[\widehat\ell_W(N)-\widehat\ell_{\gamma}(N)\right].
 \label{eq:test-statistic}
\end{equation}
For equal prior probabilities, $T>0$ assigns the event sample to the $\SUtwo$ model, and $T<0$ assigns it to the photophilic ALP case. Exact likelihood ties count as one half of a correct assignment.

At fixed event count $N$, we quote the smallest probability of choosing the correct model among both possible true models and the lifetimes in their respective domains,
\begin{equation}
 A_N=\min_{H,\,c\tau_H\in\mathcal T_H(m_a,N)}
 P_{H,c\tau_H}(\widehat H=H\mid N).
 \label{eq:accuracy}
\end{equation}
We define the required event count as
\begin{equation}
 \Nninety=\min\left\{N\,\middle|\,
 A_{N'}\geq0.90\ \text{for every tested }N'\geq N
 \right\}.
 \label{eq:N90}
\end{equation}
Requiring the accuracy to remain above 90\% at every larger tested number of events prevents one upward statistical fluctuation from setting the threshold. Wilson binomial intervals quantify the Monte Carlo uncertainty in the estimated accuracy near a crossing. This probability describes binary model assignment in pseudoexperiments in which every event is signal. Discovery and exclusion require different statistical methods.

\subsection{Testing mixtures of the interactions}
\label{sec:mixed-method}

Establishing a mixed interaction requires one test that rejects both pure limits. The null and alternative are
\begin{equation}
 H_0=H_\gamma\cup H_W,
 \qquad
 H_1=\{H_{\rm mix}:0<\xi<1\},
 \label{eq:mixed-composite-test}
\end{equation}
where $H_{\rm mix}$ allows both coefficients, their relative sign, and the lifetime to vary. Defining the best mixed and pure likelihoods by $\ell_{\rm mix}$ and $\ell_0$ gives
\begin{align}
 \ell_{\rm mix}
 &=\max_{0\leq\xi\leq1,\,s=\pm1,\,c\tau_a}
 \log\mathcal L_{\rm mix},\nonumber\\
 \ell_0
 &=\max\!\left(
 \max_{c\tau_\gamma}\log\mathcal L_\gamma,
 \max_{c\tau_W}\log\mathcal L_W
 \right),\nonumber\\
 q&=2(\ell_{\rm mix}-\ell_0).
 \label{eq:mixed-global-statistic}
\end{align}
The likelihood uses each event's energy, production angle, and proper decay distance. It is conditional on the observed count $N$, and the lifetime is fitted independently under the mixed hypothesis, $H_\gamma$, and $H_W$. For every tested $N$, each hypothesis is restricted to lifetimes that can yield that many accepted events at the SHiP exposure:
\begin{equation}
 \mathcal T_{\xi,s}(N)
 =\{c\tau_a:\mu(m_a,\xi,s,c\tau_a)\geq N\}.
 \label{eq:mixed-lifetime-domain}
\end{equation}
The restriction applies separately to the true mixture and every fitted hypothesis. The mixed scan uses the diphoton lifetime where the diphoton width controls the total width. Near destructive cancellation, subleading decay modes determine the lifetime. The critical value targets a 10\% probability of falsely rejecting either pure interaction. At each evaluated point, 1000 pseudoexperiments generated under each pure hypothesis determine the fraction of false rejection, and 1000 generated under the mixture determine the power. Appendix~\ref{sup:mixed-likelihood} gives the production weights, event density, and yield relation.

\section{Semianalytic analysis: results}
\label{sec:ideal-results}

We apply the method of \cref{sec:statistics} to the accepted decays defined in \cref{sec:geometry}, assuming perfect reconstruction. We determine the number of events needed to distinguish the pure interactions and to identify their admixtures.

\subsection{Two pure interactions}
\label{sec:pure-results}

Energy alone admits a degeneracy between the production interaction and the lifetime; the transverse decay position breaks it at the tested masses.

\subsubsection{Degeneracy of the energy distributions at different lifetimes}
\label{sec:energy-degeneracy}

A fixed longitudinal interval $z_{\min}<z<z_{\max}$ gives a simple expression for the dependence on the ALP boost. For an ALP whose trajectory crosses this interval, the decay probability is
\begin{equation}
 \begin{aligned}
 P_{\rm dec}
   &=e^{-z_{\min}/\ell_z}\left(1-e^{-\Delta z/\ell_z}\right),\\
 \ell_z&=c\tau_a\beta_a\gamma_a\cos\theta_a,
 \qquad \Delta z=z_{\max}-z_{\min}.
 \end{aligned}
 \label{eq:finite-decay-probability}
\end{equation}
Equation~\eqref{eq:finite-decay-probability} gives the longitudinal limit. The numerical calculation additionally evaluates the allowed azimuthal interval $\Delta\phi(\theta_a,z)$ set by the transverse boundaries of the trapezoidal vessel at every $(\theta_a,z)$ before applying the diphoton ECAL requirement. In the limit of a long lifetime, $\ell_z\gg z_{\max}$, the simplified expression becomes $P_{\rm dec}\simeq\Delta z/\ell_z\propto1/(\beta_a\gamma_a\cos\theta_a)$. Decay in the vessel therefore gives more relative weight to the lower energy part of each parent spectrum. Requiring both photons to reach the ECAL favors larger boosts because the photon opening angle decreases as the ALP energy rises.

\begin{figure}[t]
  \centering
  \includegraphics[width=\columnwidth]{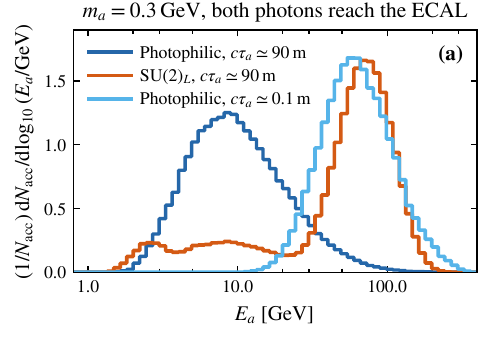}\\[-0.3em]
  \includegraphics[width=\columnwidth]{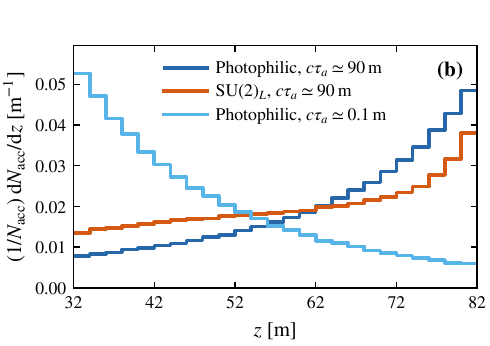}
  \caption{Probability densities $\rho_E$ in $\log_{10}(E_a/\gev)$ (upper panel) and $\rho_z$ in the longitudinal decay position $z$ (lower panel), defined in \cref{eq:accepted-distributions}, for accepted diphoton decays at $m_a=0.3~\gev$. The curves at $c\tau_a\simeq90~\m$ compare the models at a common lifetime. The pale blue photophilic curve forms the pair in \cref{eq:degenerate-lifetimes} with the electroweak curve. See the main text for the effect of the lifetime on these distributions.}
  \label{fig:energy-lifetime-degeneracy}
\end{figure}

The competition between the decay probability and the ECAL geometry acts differently on the two parent spectra. \Cref{fig:energy-lifetime-degeneracy} shows the resulting accepted energy distributions at $m_a=0.3~\gev$. At the common benchmark $c\tau_a\simeq90~\m$ introduced in \cref{sec:lifetime-domains}, the distributions remain widely separated: their medians are about $9~\gev$ for the photophilic model and $61~\gev$ for the $\SUtwo$ model.

For a short proper lifetime, the survival exponential reverses the preference for lower energies. The decay density near the entrance scales as
\begin{equation}
 \left.\frac{dP_{\rm dec}}{dz}\right|_{z\simeq z_{\min}}
 \propto
 \frac{\exp\!\left[-z_{\min}/(c\tau_a\beta_a\gamma_a\cos\theta_a)\right]}
      {\beta_a\gamma_a\cos\theta_a}.
 \label{eq:short-lifetime-decay-density}
\end{equation}
The exponential suppresses low boosts that decay upstream, and the ECAL geometry further favors decays at higher energy. For a photophilic ALP with $c\tau_\gamma\simeq0.1~\m$, the median accepted energy rises to about $62~\gev$, close to the value for the $\SUtwo$ model at long lifetime. The pair with nearly matching energy distributions is
\begin{equation}
 c\tau_{\gamma}\simeq0.1~\m,
 \qquad
 c\tau_W\simeq90~\m.
 \label{eq:degenerate-lifetimes}
\end{equation}

For each model and lifetime, we normalize the distributions by the total number $N_{\rm acc}$ of accepted decays, with the production weights included. The densities plotted in \cref{fig:energy-lifetime-degeneracy} are
\begin{equation}
 \rho_E(u)=\frac{1}{N_{\rm acc}}\frac{\diff N_{\rm acc}}{\diff u},
 \qquad
 \rho_z(z)=\frac{1}{N_{\rm acc}}\frac{\diff N_{\rm acc}}{\diff z},
 \label{eq:accepted-distributions}
\end{equation}
where $u=\log_{10}(E_a/\gev)$ and the acceptance is defined in \cref{sec:geometry}.

The lower panel of \cref{fig:energy-lifetime-degeneracy} shows why the longitudinal decay coordinate adds information. In the short-lived photophilic sample, the survival probability decreases rapidly with $z$, so accepted decays are concentrated near the upstream end of the vessel. At long lifetime, the ECAL aperture preferentially retains downstream decays of the softer photophilic ALPs because a shorter distance to the ECAL helps both photons remain inside it. The harder $\SUtwo$ sample has a broader distribution in $z$. The short-lived photophilic and long-lived electroweak samples can therefore have similar energy distributions while remaining distinct in decay position.

We quantify the separation between two normalized distributions $p$ and $q$ over common bins by their total variation distance,
\begin{equation}
 D_{\rm TV}(p,q)=\frac12\sum_i|p_i-q_i|.
 \label{eq:total-variation}
\end{equation}
It is zero for identical distributions and one for disjoint ones; for equal prior probabilities, the greatest possible accuracy from one event is $(1+D_{\rm TV})/2$. On the fine display grid, the energy distributions for the pair in \cref{eq:degenerate-lifetimes} have $D_{\rm TV}\simeq0.23$, compared with 0.70 at the common long lifetime. Production from induced photons and decays of charged kaons soften the electroweak spectrum and strengthen this similarity relative to a calculation containing only $B$ decays. Fitting the lifetime therefore substantially reduces the information carried by energy.

\subsubsection{Information from the decay position}
\label{sec:spatial-gain}

Adding decay position reduces the requirement from over one hundred accepted events to a few, as shown in \cref{tab:observable-thresholds}. The transverse coordinate provides most of the improvement; including the longitudinal coordinate gives a modest further gain.

\begin{table}[t]
\centering
\caption{Number $\Nninety$ of accepted diphoton events needed for at least 90\% correct identification, defined in \cref{eq:N90}. Results use $m_a=0.3~\gev$ and assume perfect reconstruction. The scan using only $E_a$ advances in steps of ten events.}
\label{tab:observable-thresholds}
\begin{ruledtabular}
\begin{tabular}{lc}
Observables & $\Nninety$ \\
\hline
$E_a$ & $130<\Nninety\leq140$ \\
$E_a,\langle z\rangle$ & 28 \\
$E_a,\langle\rperp\rangle$ & 8 \\
$E_a,\langle z\rangle,\langle\rperp\rangle$ & 7 \\
\end{tabular}
\end{ruledtabular}
\end{table}

The physical origin follows from \cref{eq:r-angle}. For production at the target, $\rperp$ determines the ALP polar angle once $z$ is known. The source from $B$ decays and the electromagnetic sources populate different angular distributions. Requiring both photons to reach the ECAL also affects the softer photophilic sample more strongly. The accepted $\rperp$ distribution therefore reflects both the production angle and the geometric acceptance of the photons.

For the combined observables, the requirement decreases from seven accepted decays at $m_a=0.3~\gev$ to three at $1~\gev$. \Cref{tab:joint-accuracy} gives the accuracy near these thresholds.

\begin{table*}[t!]
\centering
\caption{Minimum probability $A_N$ of correct identification, defined in \cref{eq:accuracy}, near the required number $\Nninety$ of accepted diphoton events in \cref{eq:N90}, assuming perfect reconstruction.}
\label{tab:joint-accuracy}
\begin{ruledtabular}
\begin{tabular}{lcccc}
$m_a$ & $A_{\Nninety-1}$ & $A_{\Nninety}$ & $A_{\Nninety+1}$ & $\Nninety$ \\
\hline
$0.3~\gev$ & $A_6=0.898$ & $A_7=0.919$ & $A_8=0.937$ & 7 \\
$1~\gev$ & $A_2=0.898$ & $A_3=0.945$ & $A_4=0.973$ & 3 \\
\end{tabular}
\end{ruledtabular}
\end{table*}

\subsubsection{Dependence on the energy partition}
\label{sec:generated-scope}

The number of events required depends on the energy partition. At $m_a=0.3~\gev$, a finer partition satisfying the statistical criterion of \cref{sec:probability-model} lowers the requirement from seven events to four. At $1~\gev$, broadening the first bin preserves the threshold of three events. The tested partitions thus place the requirement at a few events, with the larger variation at low mass.

Appendix~\ref{sup:pure-likelihood} gives the numerical tests of the energy partition. Appendix~\ref{sup:gaussian-resolution} examines the effect of independent, unbiased Gaussian coordinate errors while keeping the accepted events fixed.

\subsection{Mixtures of the interactions}
\label{sec:broader}
\label{sec:mixed-operators}

\begin{figure*}[t!]
 \centering
 \includegraphics[width=0.49\textwidth]{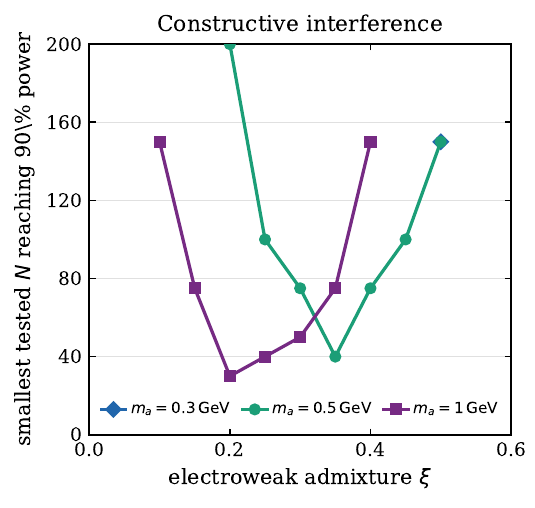}\hfill
 \includegraphics[width=0.49\textwidth]{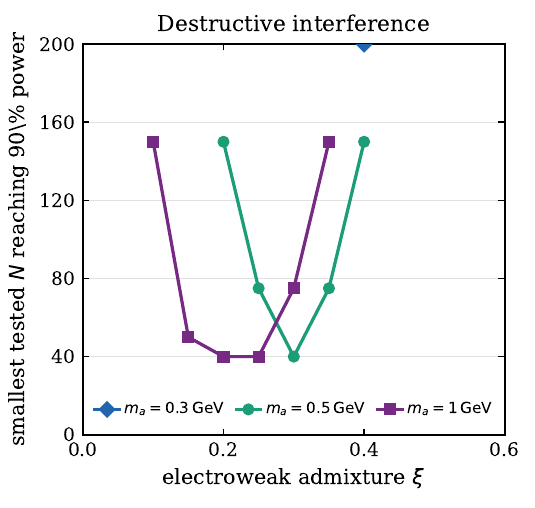}
 \caption{Number of accepted diphoton events needed to establish that both interactions contribute. Each point is the smallest tested $N$ for which the test in \cref{eq:mixed-global-statistic} rejects both pure hypotheses in at least 90\% of pseudoexperiments at the selected lifetimes, following the power criterion in \cref{sup:eq:size-power}. The admixture $\xi$ and interference signs are defined in \cref{eq:mixed-xi,eq:mixed-sign}; colors denote the mass. At $m_a=0.3~\gev$, the largest fraction of false rejection of a pure interaction is 12--13\%. Only requirements at $N\leq200$ are shown; Appendix~\ref{sup:numerics} lists the tested intervals.}
 \label{fig:mixed-operator-composite-three-mass}
\end{figure*}

The decay kinematics can reveal that both interactions contribute by excluding each pure hypothesis. \Cref{fig:mixed-operator-composite-three-mass} shows the number of accepted events needed to do so in at least 90\% of pseudoexperiments at the selected lifetimes. Favorable admixtures require several tens of events at $m_a=0.5$ and $1~\gev$, and of order one hundred at $0.3~\gev$. The scan covers lifetimes with weak separation and their neighbors; Appendix~\ref{sup:numerics} gives its settings and the intervals containing the required counts.

Near the favorable regions at $m_a=0.5$ and $1~\gev$, the sampled fractions of false rejection remain below the 10\% target, although their binomial intervals extend above it. At $0.3~\gev$, they reach 12--13\% for $N=100$--200. The curves at this mass therefore give the event counts reaching 90\% sampled power with these larger probabilities of false rejection.

For destructive interference, two isolated admixtures require separate treatment. At
\begin{equation}
 \xi_W=\frac{1}{1+2\sin^2\theta_W}\simeq0.684,
 \label{eq:mixed-cW-degeneracy}
\end{equation}
the normalized photon and flavor weights equal those of pure $C_W$ at every lifetime, so a conditional shape test cannot resolve the mixture.  At
\begin{equation}
 \xi_0=\frac{1}{1+\sin^2\theta_W}\simeq0.812,
 \label{eq:mixed-photon-cancellation}
\end{equation}
$C_\gamma^{\rm tot}=0$, and the minimal model has no diphoton signal to classify.

\section{Event requirements after detector reconstruction}
\label{sec:ecal-reco}
\begin{figure*}[t!]
  \centering
  \includegraphics[width=\textwidth]{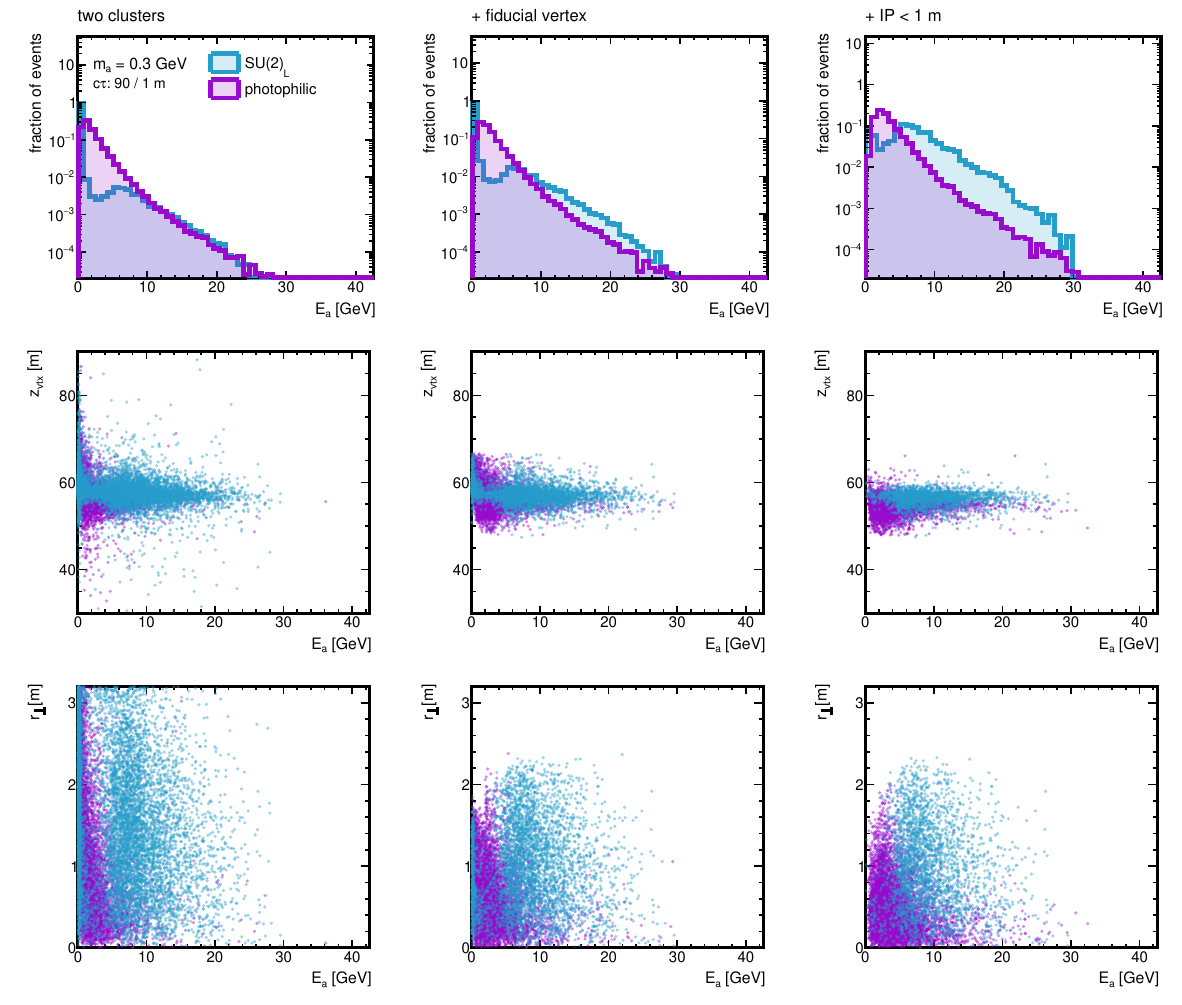}
  \caption{Distributions reconstructed by the ECAL simulation at $m_a=0.3~\gev$. Columns correspond to selections $S_1$, $S_2$, and $S_3$ in \cref{sec:reco-samples}. The top row shows the fraction of candidates in each energy bin; the other rows show reconstructed $z$ and $\rperp$ versus energy. The displayed lifetime pair, listed in \cref{tab:tvd}, minimizes the joint total variation distance defined in \cref{eq:total-variation} after $S_3$ among the simulated pairs.}
  \label{fig:reco-shapes-0p3}
\end{figure*}

We compare the two pure interactions using diphoton candidates after calorimeter reconstruction and selection. Generated diphoton decays are propagated through the full \textsc{Geant4} description of the SHiP electromagnetic calorimeter and reconstructed with the procedure of Ref.~\cite{Climescu:2026esp}. Appendix~\ref{sup:detbase-reco} gives the simulation sizes.

Each candidate has two reconstructed showers with energies $E_{\gamma1}$ and $E_{\gamma2}$. Their axes are obtained from the transverse and longitudinal cluster profiles in successive calorimeter layers of varying granularity. The reconstructed ALP energy is $E_a^{\rm rec}=E_{\gamma1}+E_{\gamma2}$, and its decay vertex is the point of closest approach of the two axes. The coordinates $z_{\rm rec}$ and $r_\perp^{\rm rec}=\sqrt{x_{\rm rec}^2+y_{\rm rec}^2}$ give the longitudinal position and distance from the beam axis. The reconstructed ALP direction is the energy-weighted sum of the two photon directions. The impact parameter used in selection $S_3$ is the three-dimensional distance from the nominal target point to the line through the reconstructed vertex along that direction. Since an ALP produced at the target has zero impact parameter, its reconstructed value measures the quality of the vertex and direction reconstruction.

\subsection{Samples and selections}
\label{sec:reco-samples}
\begin{figure*}[t!]
  \centering
  \includegraphics[width=\textwidth]{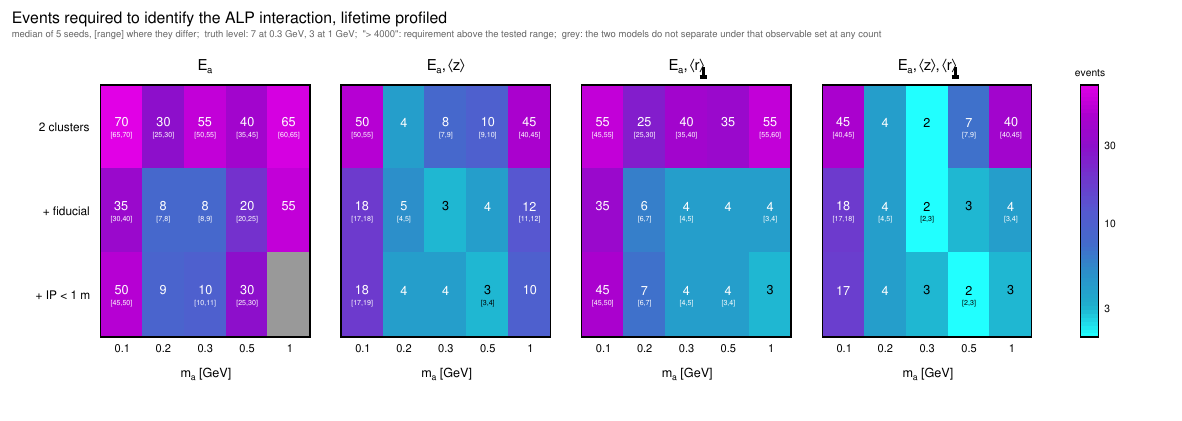}
   \caption{Number $\Nninety$ of reconstructed diphoton events needed to distinguish the two pure interactions with at least 90\% correct identification, defined in \cref{eq:N90}. Rows correspond to selections $S_1$, $S_2$, and $S_3$ in \cref{sec:reco-samples}; columns specify the observables used in the likelihood. Each cell gives the median over five random seeds, with the full range in brackets. Gray cells did not reach $A_N=0.90$ over the tested range of $N$; at $m_a=1~\gev$, the range ends where no simulated lifetime satisfies \cref{eq:reco-lifetime-domain}.}
  \label{fig:profiled-grid}
\end{figure*}

Both models are simulated at $m_a=0.1$, $0.2$, $0.3$, $0.5$, and $1~\gev$, with five to nine lifetimes per mass. The photophilic events include primary and cascade Primakoff production. The $\SUtwo$ events include these photon sources through the induced coupling, inclusive decays of $B$ mesons, and charged-kaon decays wherever $m_a<m_K-m_\pi$. All components use the absolute \textsc{SensCalc} production yields and the \textsc{EventCalc-SHiP} propagation of \cref{sec:eventcalc}. The target production point and fiducial geometry are those of \cref{sec:geometry}; the full detector simulation follows the photon showers through the calorimeter.

We apply three nested selections based on Ref.~\cite{Climescu:2026esp}:
\begin{enumerate}
\item $S_1$: two reconstructed clusters in the calorimeter;
\item $S_2$: the $S_1$ candidates with a reconstructed decay vertex inside the fiducial volume, reduced by the local vertex resolution;
\item $S_3$: the $S_2$ candidates with an impact parameter relative to the target below $1~\m$.
\end{enumerate}

The fiducial volume defined in $S_2$ sees each vessel boundary moved inward by one corresponding local standard deviation of the vertex resolution; a full width $w$ becomes $w-2\sigma$. The local resolution map is parameterized by the two photon energies and their opening angle and evaluated event by event. Events from both models and all simulated lifetimes are pooled when constructing the map. At $m_a=0.3~\gev$, the inclusive Cartesian resolutions are approximately $\sigma_x=0.38~\m$, $\sigma_y=0.52~\m$, and $\sigma_z=12~\m$. The intersection of nearly parallel shower axes constrains the transverse coordinates more accurately than the longitudinal coordinate. We obtain the response of the nonnegative radial coordinate $\rperp$ directly from the eventwise reconstruction; the quoted resolutions summarize the Cartesian coordinates.

Let $D$ denote the accepted decays defined in \cref{sec:geometry}, weighted by the absolute production yields of \cref{sec:eventcalc}. For model $H$, the efficiency of selection $S_j$ is the fraction of these decays that survives the selection. With event weights $w_{Hk}$, it is
\begin{equation}
 \varepsilon_H^{S_j}(m_a,c\tau_a)
 =\frac{\sum_{k\in S_j}w_{Hk}}{\sum_{k\in D}w_{Hk}}.
 \label{eq:reco-efficiency}
\end{equation}
The lifetime domain of \cref{eq:lifetime-domain} has a direct detector analog. Writing $\mathcal G_H(m_a)$ for the set of lifetimes simulated for model $H$ at that mass, the profiled lifetimes after selection $S_j$ are

\begin{equation}
 \begin{aligned}
 \mathcal T_H^{S_j}(m_a,N)
 ={}&\bigl\{c\tau_a\in\mathcal G_H(m_a):\\
 &\mu_H^{S_j}(m_a,c\tau_a)\geq N\bigr\}
 \setminus\mathcal E_H(m_a),
 \end{aligned}
 \label{eq:reco-lifetime-domain}
\end{equation}

where $\mu_H^{S_j}=\varepsilon_H^{S_j}\,\mu_H^{\rm acc}$ is the mean number of candidates surviving that selection. We apply this condition at every tested $N$ and require at least one hundred simulated candidates to survive $S_j$ at each fitted lifetime. The likelihood in \cref{eq:profiled-likelihood} then uses the selected candidates and the domain in \cref{eq:reco-lifetime-domain}.

\Cref{fig:reco-shapes-0p3} shows how the reconstructed distributions change across the three selections at $m_a=0.3~\gev$.

\subsection{Separation of the reconstructed distributions}
\label{sec:reco-separation}
\begin{figure*}[t!]
  \centering
  \includegraphics[width=\textwidth]{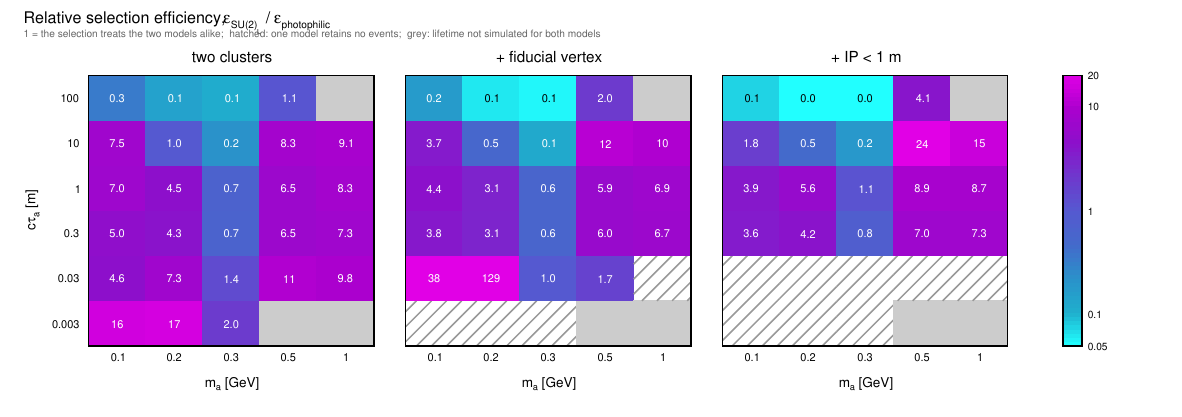}
    \caption{Ratio $\varepsilon_{\SUtwo}^{S_j}/\varepsilon_\gamma^{S_j}$ of the efficiencies defined in \cref{eq:reco-efficiency}, at each mass and lifetime simulated for both models. Panels show selections $S_1$, $S_2$, and $S_3$. Hatching denotes zero selected events for one model; gray denotes that the models have no common simulated lifetime. See the main text for the effect of the selections.}

  \label{fig:efficiency-grid}
\end{figure*}

The total variation distance gives a diagnostic of the separation between two fixed reconstructed distributions. \Cref{tab:tvd} gives this distance for the joint $(E_a,z,\rperp)$ histogram after $S_3$, minimized over the simulated pairs of lifetimes. This minimum identifies the closest fixed pair on the detector grid. The profiled classifier tests every allowed lifetime for every pseudoexperiment, so its event requirement is a separate quantity.
 
\begin{table*}[t!]
\centering
\caption{Smallest total variation distance $D_{\rm TV}$ defined in \cref{eq:total-variation}, between reconstructed joint distributions in $(E_a,z,\rperp)$ after selection $S_3$. At each mass, the minimum is taken over all simulated pairs of lifetimes. For equal prior probabilities, the greatest possible accuracy from one event is $(1+D_{\rm TV})/2$, as shown in the last column.}
\label{tab:tvd}
\begin{ruledtabular}
\begin{tabular*}{\textwidth}{@{\extracolsep{\fill}}lccc@{}}
$m_a$ [$\gev$] & lifetimes at minimum & $D_{\rm TV}$ & greatest accuracy from one event \\
\hline
0.1 & $\gamma$ $10~\m$  vs.\ $W$ $100~\m$  & 0.10 & 0.55 \\
0.2 & $\gamma$ $1~\m$   vs.\ $W$ $1000~\m$ & 0.41 & 0.71 \\
0.3 & $\gamma$ $1~\m$   vs.\ $W$ $90~\m$   & 0.55 & 0.78 \\
0.5 & $\gamma$ $1~\m$   vs.\ $W$ $1000~\m$ & 0.64 & 0.82 \\
1.0 & $\gamma$ $0.1~\m$ vs.\ $W$ $0.3~\m$  & 0.59 & 0.80 \\
\end{tabular*}
\end{ruledtabular}
\end{table*}
 
For a fixed pair of lifetimes, the separation is smallest at the lowest mass studied. There, the approximate opening angle $2m_a/E_a$ of a symmetric decay is smallest, and the vertex follows from two nearly parallel shower axes. The separation is larger at intermediate masses and decreases again when the reconstructed energy spectra converge near the upper end of the range. The observable study below shows that $\rperp$ carries most of the spatial discrimination, consistent with \cref{sec:spatial-gain}.
 
\subsection{Event requirement}
\label{sec:reco-n90}
\begin{figure*}[t!]
\centering
\includegraphics[width=0.49\textwidth]{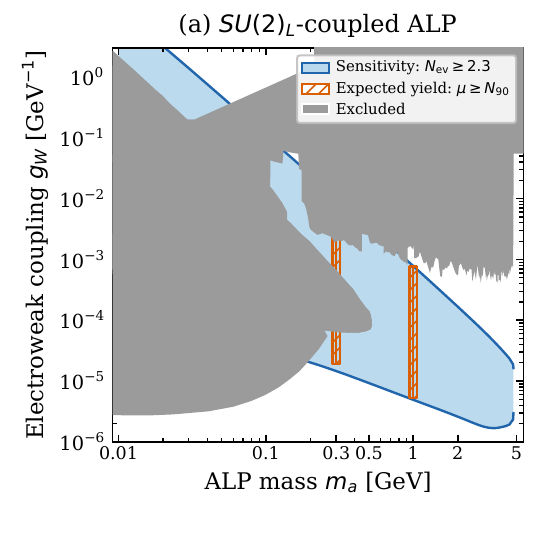}\hfill
\includegraphics[width=0.49\textwidth]{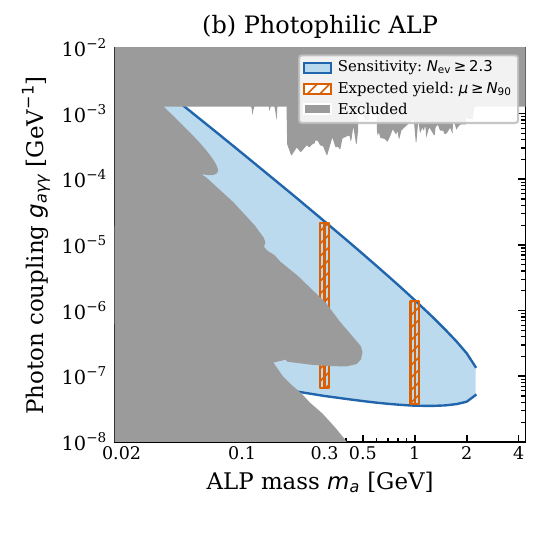}
\caption{Regions in which SHiP can observe a diphoton signal (blue) and identify its production interaction assuming perfect reconstruction (hatching). The left panel assumes a pure electroweak interaction; the right panel assumes a pure photophilic interaction. The blue boundary satisfies \cref{eq:sensitivity}; the hatched regions satisfy \cref{eq:identification-domain} with $\Nninety=7$ at $m_a=0.3~\gev$ and $\Nninety=3$ at $1~\gev$. Their widths serve only to display these two discrete masses. Gray regions are excluded by existing searches~\cite{deBlas:2025gyz,Ferber:2022rsf}. See the main text for the exposure, acceptance, and branching fraction.}
\label{fig:summary-reach}
\end{figure*}

\Cref{fig:profiled-grid} shows the required numbers of reconstructed candidates for each mass, selection, and choice of observables. Appendix~\ref{sup:detbase-reco} lists the values for the combined observables in \cref{tab:n90reco}. We repeat the pseudoexperiments with five random seeds and quote the median, with the full range in brackets.

For each model and lifetime, the energy probabilities, conditional means, and covariances in \cref{eq:likelihood} are estimated from the candidates surviving $S_j$, using their reconstructed energies and coordinates. Pseudoexperiments draw energy bins and Gaussian spatial means as described in \cref{sec:probability-model}, with the moments in \cref{sup:eq:conditional-moments}. Generated coordinates enter only through the residuals used to build the resolution map for $S_2$.

After $S_3$, the lowest tested mass requires 17 reconstructed candidates, while the other masses require two to four. The increase at low mass follows the poorer vertexing associated with smaller diphoton opening angles.

The requirements in \cref{tab:joint-accuracy} and \cref{fig:profiled-grid} refer to accepted decays and selected reconstructed candidates, respectively. Selection changes the relative populations of the two models: \cref{fig:efficiency-grid} shows efficiency ratios departing from unity by an order of magnitude and reversing between low and intermediate mass. The largest ratios occur at short lifetimes: at $m_a=0.2~\gev$ and $c\tau_a=0.03~\m$, the photophilic efficiency after $S_2$ is below one percent of its value after $S_1$. At $m_a=0.3~\gev$, the overlap of the reconstructed energy spectra is $0.47$ among candidates rejected by $S_1$, $0.41$ among those retained, and $0.39$ after $S_3$. The requirement per reconstructed candidate therefore describes distributions reshaped by selection as well as detector response. The efficiencies in \cref{eq:reco-efficiency} relate the numbers of accepted and reconstructed events.

Energy alone fails to reach the target in parts of the grid. At $m_a=1~\gev$ after $S_3$, the accuracy obtained from energy remains below 90\% through the largest tested count, $N=45$. The reconstructed energy distributions remain close for the limiting pair of lifetimes, preserving the degeneracy between production and lifetime seen with generated quantities.

\section{Observation and identification reach}
\label{sec:reach}

To determine the reach for identifying either pure interaction, we combine the requirements from the semianalytic study in \cref{sec:pure-results}, assuming perfect reconstruction, with the mean accepted yield at the exposure in \cref{eq:exposure}. For a true model $H$, we define
\begin{equation}
 \widetilde{\mathcal{D}}_H^{\rm acc}=
 \left\{(m_a,g_H)\,\middle|\,
 \mu_H^{\rm acc}(m_a,g_H)\geq
 N_{90}^{\rm acc}(m_a)
 \right\},
 \label{eq:identification-domain}
\end{equation}
where $\mu_H^{\rm acc}$ is the \textsc{EventCalc-SHiP} mean accepted yield and $N_{90}^{\rm acc}$ follows from \cref{tab:joint-accuracy}. When the mean equals the required integer count, the Poisson probability of observing at least that many events is about 55--58\% for the two examples below. Thus \cref{eq:identification-domain} defines a boundary based on the expected yield.

\Cref{fig:summary-reach} compares this region with the sensitivity boundary in \cref{eq:sensitivity}. The accepted yield is evaluated over the full mass range shown, while the identification requirements are calculated at the two masses in \cref{tab:joint-accuracy}. The branching fraction follows the convention in \cref{sec:interactions}.

At the tested masses, the identification region for each pure interaction spans nearly the entire coupling range within SHiP's projected exclusion sensitivity. The few events needed for discrimination restrict this region only near the sensitivity boundaries, where the expected yield is smallest.

\section{Conclusions}
\label{sec:conclusions}

If SHiP discovers a new particle, identifying the interaction responsible for its production will be essential to understanding its nature. SHiP cannot observe the particle production process, as it occurs inside its thick target. When different interactions produce the same visible final state, their origin must therefore be inferred from the particle decay kinematics. We have explored this possibility for ALPs with couplings to the $U(1)_Y$ and $SU(2)_L$ gauge fields and a dominant diphoton decay. The ALP mass can be reconstructed from the energies and opening angle of the two photons, while the unknown lifetime is fitted independently under each production hypothesis.

We first performed the study assuming perfect event reconstruction (\cref{sec:ideal-results}) and found that adjusting the lifetime can make the energy distributions from different production interactions nearly identical. Combining the energy with the longitudinal and transverse decay positions resolves this ambiguity, with the transverse position providing the largest improvement in discrimination.

The full simulation of the SHiP electromagnetic calorimeter, including photon showers and diphoton reconstruction (\cref{sec:ecal-reco}), provides, to our knowledge, the first demonstration at detector level of SHiP's ability to identify the production interaction after discovery. With the selections considered, only $2$--$4$ reconstructed diphoton events are needed to distinguish the pure $U(1)_Y$ and $SU(2)_L$ interactions for ALP masses between $0.2$ and $1~\mathrm{GeV}$, with at least $90\%$ correct identification (\cref{fig:profiled-grid}).

We also explored whether the decay kinematics can reveal that both interactions contribute. Their relative magnitude and sign affect the production distributions and the diphoton decay amplitude. The study assuming perfect reconstruction indicates that several tens to $\mathcal{O}(10^2)$ events can distinguish favorable admixtures from both pure interactions (\cref{fig:mixed-operator-composite-three-mass}). The number of events needed depends strongly on the couplings. Certain mixtures produce the same kinematic distributions as a pure interaction and remain indistinguishable through these observables.

At the tested masses, the small number of accepted decays needed to distinguish the pure interactions gives an identification reach close to SHiP's projected sensitivity (\cref{fig:summary-reach}). Exploiting this potential with reconstructed diphoton events requires efficient retention of both photon showers and accurate determination of the ALP energy and transverse decay position.

The same logic can be applied to more complex signals, including fully reconstructible final states and partially reconstructible ones for which only a subset of the decay products is used. The comparison would use the momenta of the reconstructed particles and the available information on the decay position, with the unobserved kinematics included in the signal model.

\begin{acknowledgments}
We acknowledge the use of GPT-Sol 5.6 and GPT-Astra 6 for assistance with code development and manuscript preparation. The authors take full responsibility for the content.
M.F.M. thanks CERN for its hospitality during his participation in the CERN Summer Student Programme.
\end{acknowledgments}

\section*{Conflict of Interest Statement}
Some of the authors (MC, MO) are also members of the SHiP collaboration. The present manuscript solely reflects the authors' views and not those of the SHiP collaboration.

\bibliography{references}

\newpage
\clearpage 
\appendix

\section{Electroweak ALP production}
\label{sup:production}

The $\SUtwo$ interaction induces the flavor-changing coupling
\begin{equation}
 g_{bq}^{a}=
 \frac{3g_2^2}{16\pi^2}\frac{\alpha_2}{4\pi}g_W
 \sum_{i=u,c,t}V_{ib}V_{iq}^{*}G(x_i),
 \qquad q=s,d,
 \label{sup:eq:gbq}
\end{equation}
where
\begin{equation}
 x_i=\frac{m_i^2}{m_W^2},
 \qquad
 G(x)=\frac{x[1+x(\log x-1)]}{(1-x)^2}.
 \label{sup:eq:loop-function}
\end{equation}
The term from the top quark dominates. For a pseudoscalar final state $P=K,\pi$,
\begin{multline}
 \Gamma(B^+\to P^+a)=
 \frac{m_B^3}{64\pi}|g_{bq}^{a}|^2
 \left[f_0^{B\to P}(m_a^2)\right]^2\\
 \times\lambda_{Pa}^{1/2}
 \left(1-\frac{m_P^2}{m_B^2}\right)^2,
 \label{sup:eq:BPa-width}
\end{multline}
with
\begin{equation}
 \lambda_{Pa}=
 \left[1-\frac{(m_P+m_a)^2}{m_B^2}\right]
 \left[1-\frac{(m_P-m_a)^2}{m_B^2}\right].
 \label{sup:eq:lambda-Pa}
\end{equation}
The inclusive production probability is
\begin{multline}
 P_{a,\rm prod}=2\chi_{b\bar b}
 \left(f_{b\to B^+}+r_{B^0/B^+}f_{b\to B^0}\right)\\
 \times\sum_X\br(B^+\to X^+a),
 \label{sup:eq:inclusive-B}
\end{multline}
where $\chi_{b\bar b}=2.7\times10^{-7}$ is the probability of producing a $b\bar b$ pair per proton on target~\cite{Boiarska:2019jym}, $f_{b\to B^+}=0.417$ and $f_{b\to B^0}=0.418$ are fragmentation fractions~\cite{Bondarenko:2019yob}, and $r_{B^0/B^+}\equiv\br(B^0\to X a)/\br(B^+\to X^+a)=0.93$ is the assumed ratio of inclusive ALP branching fractions~\cite{Boiarska:2019jym}.

The derivative interaction obeys
\begin{equation}
 q_{\mu}\overline q\gamma^{\mu}P_Lb
 =m_b\overline qP_Rb-m_q\overline qP_Lb.
 \label{sup:eq:EOM-current}
\end{equation}
After neglecting $m_q/m_b$, its hadronic matrix element is proportional to the same leading current used for a scalar mixed with the Higgs boson~\cite{Boiarska:2019jym}. Overall Wilson coefficients cancel from the exclusive ratios $\br(B\to X_sa)/\br(B\to Ka)$. Corrections from the strange quark mass enter at order $m_s/m_b\simeq2\%$ in the amplitude.

Photon production induced by the electroweak interaction follows
\begin{equation}
 \frac{P_{\rm photonic}}{g_W^2}
 =\left(\frac{\alpha_{\rm em}}{\pi}\right)^2
  \frac{P_{\rm photonic}}{\gagg^2}.
 \label{sup:eq:production-conversion}
\end{equation}
For charged kaons,
\begin{equation}
 \frac{P_{K\to\pi a}}{g_W^2}
 =0.36\,\frac{\br(K^\pm\to\pi^\pm a)}{g_W^2},
 \label{sup:eq:kaon-production-coefficient}
\end{equation}
where 0.36 is the number of charged kaons of both signs that decay in flight per proton on target. \textsc{EventCalc-SHiP} generates the isotropic two-body decay and combines the distributions from kaons, $B$ mesons, and photons with their absolute production yields.

\section{Likelihood for the two pure interactions}
\label{sup:pure-likelihood}

The likelihood and lifetime domain are defined in \cref{eq:likelihood,eq:lifetime-domain}. For $N$ independent events with energy bins $\boldsymbol i=(i_1,\ldots,i_N)$, the conditional mean and covariance of $\overline{\boldsymbol f}$ are
\begin{align}
 \overline{\boldsymbol\mu}_H(c\tau_H\mid\boldsymbol i)
 &=\frac{1}{N}\sum_{k=1}^N\boldsymbol\mu_{H,i_k}(c\tau_H),\nonumber\\
 \boldsymbol V_H(c\tau_H\mid\boldsymbol i)
 &=\frac{1}{N^2}\sum_{k=1}^N\boldsymbol\Sigma_{H,i_k}(c\tau_H).
 \label{sup:eq:conditional-moments}
\end{align}
These moments determine the Gaussian used both to generate pseudoexperiments and to evaluate the likelihood. Classification and its accuracy follow \cref{eq:test-statistic,eq:accuracy}.

At $m_a=0.3~\gev$, each source uses $10^7$ interpolation points and $10^6$ resampled parent particles. At $1~\gev$, the corresponding numbers are $10^6$ and $10^5$. Five statistically independent batches of pseudoexperiments are pooled for each true lifetime and each $N$. The joint calculation with broad bins uses at least 5000 pseudoexperiments per batch over the full allowed lifetime range, with larger samples at the limiting points.

At $m_a=0.3~\gev$, the calculation with broad energy bins and energy alone gives $A_{130}=0.8966$ and $A_{140}=0.9055$, hence $130<N_{90}\leq140$ when $N$ is increased in steps of ten. For energy plus the mean longitudinal position, the boundary values over the full lifetime domain are $A_{27}=0.8983$ and $A_{28}=0.9019$, giving $N_{90}=28$ on the tested range. Energy plus the mean transverse position gives $A_7=0.8915$ and $A_8=0.9127$, hence $N_{90}=8$. With both coordinates, the values in \cref{tab:joint-accuracy} give $N_{90}=7$ on the tested range through 20 events.

At $0.3~\gev$, the broad first bin is $E_a\in[0.3,101.9]~\gev$. A finer supported partition gives $A_3=0.8736$ and $A_4=0.9158$ after the same lifetime restriction, hence $N_{90}=4$ on the tested range. At $1~\gev$, the adaptive calculation and the calculation with broad bins both give $N_{90}=3$. With more pseudoexperiments, the boundary values are $A_2=0.8975$ with a 95\% interval of $[0.8956,0.8994]$, and $A_3=0.9454$ with a 95\% interval of $[0.9434,0.9474]$.

\section{Conditional likelihood for mixed interactions}
\label{sup:mixed-likelihood}

For production at the target, each accepted event is represented by
\begin{align}
 \theta_{a,n}&=\arctan\!\left(\frac{r_{\perp,n}}{z_n}\right),\nonumber\\
 \ell_{0,n}&=\frac{m_a}{p_{a,n}}
 \sqrt{z_n^2+r_{\perp,n}^2},
 \qquad p_{a,n}=\sqrt{E_{a,n}^2-m_a^2}.
 \label{sup:eq:eventwise-coordinates}
\end{align}
The variable $\ell_0$ is the proper decay length, equal to $c$ times the elapsed proper time. A common finite partition of $(E_a,\theta_a,\ell_0)$ retains the correlations between energy and position and the distinct populations from photon and flavor production.

Let $h_\gamma(b\mid c\tau_a)$ be the accepted weight from photon production per $g_{a\gamma\gamma}^2$ in cell $b$, and let $h_W(b\mid c\tau_a)$ be the corresponding weight from flavor production per $g_W^2$. The latter includes inclusive decays of $B$ mesons and, at $m_a=0.3~\gev$, decays of charged kaons. With
\begin{equation}
 d_s(\xi)=s(1-\xi)+\sin^2\theta_W\xi,
\end{equation}
the normalized event density is
\begin{align}
 w_b(\xi,s,c\tau_a)
 &=16d_s^2(\xi)h_\gamma(b\mid c\tau_a)\nonumber\\
 &\quad+\left(\frac{4\pi}{\alpha_2}\right)^2
 \xi^2h_W(b\mid c\tau_a),\nonumber\\
 p_b(\xi,s,c\tau_a)&=
 \frac{w_b(\xi,s,c\tau_a)}{\sum_{b'}w_{b'}(\xi,s,c\tau_a)}.
 \label{sup:eq:mixed-density}
\end{align}
The likelihood conditional on $N$ is
\begin{equation}
 \log\mathcal L_{\rm evt}(\xi,s,c\tau_a)
 =\sum_{n=1}^{N}\log p_{b_n}(\xi,s,c\tau_a).
 \label{sup:eq:mixed-likelihood}
\end{equation}

Within the benchmark dominated by the diphoton width, the proper lifetime fixes the common coefficient scale $\Lambda=|C_\gamma^{\rm dir}|+|C_W|$ through
\begin{equation}
 c\tau_a(\Lambda,\xi,s)
 =\frac{4\pi\hbar c}{m_a^3\Lambda^2d_s^2(\xi)}.
 \label{sup:eq:lifetime-scale}
\end{equation}
If $\widehat A_\gamma(c\tau_a)$ and $\widehat A_W(c\tau_a)$ are the accepted probabilities per proton on target and per squared photon and electroweak coupling, respectively, the expected accepted count is
\begin{multline}
 \mu(\Lambda,\xi,s)=N_{\rm POT}\Lambda^2
 \bigg[16d_s^2(\xi)\widehat A_\gamma(c\tau_a)\\
 +\left(\frac{4\pi}{\alpha_2}\right)^2\xi^2
 \widehat A_W(c\tau_a)\bigg].
 \label{sup:eq:mixed-yield}
\end{multline}
This yield determines the lifetime domain in \cref{eq:mixed-lifetime-domain}.

For each $N$, a common critical value satisfies
\begin{align}
 \sup_{H\in H_0,\,c\tau_H\in\mathcal T_H(N)}
 \Pr_{H,c\tau_H}(q>q_{\rm crit})&\leq0.10,\nonumber\\
 \inf_{c\tau_a\in\mathcal T_{\xi,s}(N)}
 \Pr_{\xi,s,c\tau_a}(q>q_{\rm crit})&\geq0.90.
 \label{sup:eq:size-power}
\end{align}
The first line controls false rejection of either pure interaction. The second defines the formal power target.

\section{Numerical scan at three masses}
\label{sup:numerics}

The calculation uses 80 proper lifetimes from $0.005$ to $700~\m$. The tested counts through 100 are $N=2,3,5,7,10,15,20,25,30,40,50,75,100$. We also test $N=150,200,300,500$. Independent source samples determine the likelihood representation and test its probability of false rejection and its power. The event partitions contain 176 cells at $m_a=0.3~\gev$, 774 at $0.5~\gev$, and 339 at $1~\gev$. Each evaluated point uses 1000 pseudoexperiments. The scan tests selected lifetimes at which the separation is smallest and their neighbors; \cref{sup:tab:mixed-brackets} lists the resulting intervals in $N$.

\begin{table*}[t!]
\centering
\caption{Intervals in the number $N$ of accepted diphoton events needed to identify both interactions. The upper endpoint is the first tested $N$ for which the test in \cref{eq:mixed-global-statistic} rejects both pure hypotheses in at least 90\% of pseudoexperiments at the selected lifetimes; the lower endpoint is the preceding tested value. Power and the lifetime domain are defined in \cref{sup:eq:size-power,eq:mixed-lifetime-domain}. Only intervals with upper endpoints at $N\leq200$ are listed.}
\label{sup:tab:mixed-brackets}
\begin{ruledtabular}
\begin{tabular}{cccc}
$m_a$ [$\gev$] & interference & $\xi$ & interval in $N$ \\
\hline
0.3 & constructive & 0.50 & $(100,150]$ \\
0.3 & destructive  & 0.40 & $(150,200]$ \\
\hline
0.5 & constructive & 0.20 & $(150,200]$ \\
0.5 & constructive & 0.25 & $(75,100]$ \\
0.5 & constructive & 0.30 & $(50,75]$ \\
0.5 & constructive & 0.35 & $(30,40]$ \\
0.5 & constructive & 0.40 & $(50,75]$ \\
0.5 & constructive & 0.45 & $(75,100]$ \\
0.5 & constructive & 0.50 & $(100,150]$ \\
0.5 & destructive  & 0.20 & $(100,150]$ \\
0.5 & destructive  & 0.25 & $(50,75]$ \\
0.5 & destructive  & 0.30 & $(30,40]$ \\
0.5 & destructive  & 0.35 & $(50,75]$ \\
0.5 & destructive  & 0.40 & $(100,150]$ \\
\hline
1.0 & constructive & 0.10 & $(100,150]$ \\
1.0 & constructive & 0.15 & $(50,75]$ \\
1.0 & constructive & 0.20 & $(25,30]$ \\
1.0 & constructive & 0.25 & $(30,40]$ \\
1.0 & constructive & 0.30 & $(40,50]$ \\
1.0 & constructive & 0.35 & $(50,75]$ \\
1.0 & constructive & 0.40 & $(100,150]$ \\
1.0 & destructive  & 0.10 & $(100,150]$ \\
1.0 & destructive  & 0.15 & $(40,50]$ \\
1.0 & destructive  & 0.20 & $(30,40]$ \\
1.0 & destructive  & 0.25 & $(30,40]$ \\
1.0 & destructive  & 0.30 & $(50,75]$ \\
1.0 & destructive  & 0.35 & $(100,150]$ \\
\end{tabular}
\end{ruledtabular}
\end{table*}

At $m_a=0.3~\gev$, the contribution from flavor production includes inclusive decays of $B$ mesons and four independent samples of charged kaons containing $2^{22}$ Sobol points each. These samples give accepted kaon yields consistent within about 1\%. Near the smallest displayed requirements, the largest observed fractions of false rejection in 1000 pseudoexperiments generated under a pure hypothesis are 0.093 at $m_a=0.5~\gev$ and 0.096 at $1~\gev$; their binomial intervals extend above 0.10. At $0.3~\gev$, the observed values are 0.123, 0.128, and 0.120 for $N=100$, 150, and 200. The curves at $0.3~\gev$ therefore show the sampled crossings at which the power reaches 90\%.

\section{Illustrative Gaussian smearing of decay coordinates}
\label{sup:gaussian-resolution}

The likelihood based on generated quantities can be modified to illustrate the sensitivity to unbiased coordinate errors while keeping the population of accepted events fixed. We define
\begin{equation}
 z_{\rm rec}=z+\delta_z,
 \qquad
 \rperp^{\rm rec}=\rperp+\delta_r,
 \label{sup:eq:gaussian-coordinates}
\end{equation}
where $\delta_z$ and $\delta_r$ are independent Gaussian variables with zero mean and standard deviations $\sigma_z$ and $\sigma_{r_\perp}$. In energy bin $i$, the conditional covariance becomes
\begin{equation}
 \boldsymbol\Sigma^{\rm G}_{H,i}
 =\boldsymbol\Sigma_{H,i}
 +\begin{pmatrix}
   \sigma_z^2 & 0\\
   0 & \sigma_{r_\perp}^2
  \end{pmatrix}.
 \label{sup:eq:gaussian-resolution}
\end{equation}
The same covariance is used to generate pseudoexperiments and to evaluate every fitted lifetime. The energy probabilities, conditional means, accepted population, and expected yield are unchanged. For $N$ events, the added covariance of the sample mean is $\operatorname{diag}(\sigma_z^2,\sigma_{r_\perp}^2)/N$.

The lifetime domain remains that of \cref{eq:lifetime-domain}. The full detector response and reconstruction selections are treated in \cref{sec:ecal-reco}.

\section{Detector-based reconstruction}
\label{sup:detbase-reco}
\begin{table*}[t!]
\centering
\caption{Number $\Nninety$ of reconstructed diphoton events needed for at least 90\% correct identification, defined in \cref{eq:N90}, using $(E_a,\langle z\rangle,\langle\rperp\rangle)$ and the lifetime domain in \cref{eq:reco-lifetime-domain}. Columns give $m_a$ in GeV. Entries give the median over five seeds; brackets give the full range.}
\label{tab:n90reco}
\begin{ruledtabular}
\begin{tabular*}{\textwidth}{@{\extracolsep{\fill}}lccccc@{}}
selection & 0.1 & 0.2 & 0.3 & 0.5 & 1.0 \\
\hline
two clusters       & 45 [40,45] & 4       & 2       & 7 [7,9] & 40 [40,45] \\
+ fiducial vertex  & 18 [17,18] & 4 [4,5] & 2 [2,3] & 3       & 4 [3,4] \\
+ impact parameter & 17         & 4       & 3       & 2 [2,3] & 3 \\
\end{tabular*}
\end{ruledtabular}
\end{table*}

Each generator sample contains of order $10^{5}$ decays, which are propagated through the \textsc{Geant4} simulation in slices of $2000$ events. After reconstruction, each sample yields between $6.5\times10^{4}$ and $9.8\times10^{4}$ candidates for which the reconstruction converged. The effective weighted statistics after $S_3$ vary strongly with lifetime. The simulated masses and lifetime coverage are described in \cref{sec:reco-samples}. The energy partition is merged until every bin holds at least one hundred effective entries, giving 19--22 bins at the masses studied.
 
The five seeds in \cref{tab:n90reco} vary the pseudoexperiments while keeping the simulated detector events fixed. The bracketed ranges therefore quantify Monte Carlo fluctuations in the classification scan. The separate requirement of one hundred effective entries per energy bin limits the statistical error on each estimated bin probability to about ten percent. The total variation distances in \cref{tab:tvd} also bound the accuracy from one event by $(1+D_{\rm TV})/2$; the calculated accuracies satisfy this bound.

\end{document}